\documentclass[12pt,amsmath,amssymb,]{revtex4}
\usepackage{amsfonts}
\usepackage{amsmath}
\usepackage{amssymb}
\usepackage{array}
\usepackage{bm}
\usepackage{braket}
\usepackage{breqn}
\usepackage{color}
\usepackage{dcolumn}
\usepackage{epsfig}
\usepackage{epsf}
\usepackage{epstopdf}
\usepackage{float}
\usepackage{graphicx}
\usepackage{graphics}
\usepackage{harpoon}
\usepackage{indentfirst}
\usepackage{mathrsfs}
\usepackage{multirow}
\usepackage{pifont}
\usepackage{xcolor}

\newcommand {\sla}[1]{ #1 \!\!\!/}

\newcommand{\beq}{\begin{eqnarray}}
\newcommand{\eeq}{\end{eqnarray}}

\begin{document}

\title{Two-Photon-Exchange effects in $ep\rightarrow en\pi^+$ and their corrections to separated cross sections $\sigma_{\textrm{L,T,LT,TT}}$ at small $-t$}
\author{
Hui-Yun Cao, Hai-Qing Zhou \protect\footnotemark[1] \protect\footnotetext[1]{E-mail: zhouhq@seu.edu.cn} \\
School of Physics,
Southeast University, Nanjing 211189, China}
\date{\today}

\begin{abstract}
In this work, the two-photon-exchange (TPE) effects in $ep\rightarrow en\pi^+$ at small $-t$  are discussed within a hadronic model. Under the pion dominance approximation the TPE contribution to the amplitude can be described by a scalar function in the limit $m_e\rightarrow 0$. The TPE contributions to the amplitude and the unpolarized differential cross section are both estimated when only the elastic intermediate state is considered. We find that the TPE corrections to the unpolarized differential cross section are about from $-4\%$ to $-20\%$ at $Q^2=1$--$ 1.6$ GeV$^2$. After considering the TPE corrections to the experimental data sets of unpolarized differential cross section, we analyze the TPE corrections to the separated cross sections $\sigma_{\textrm{L,T,LT,TT}}$.  We find that the TPE corrections (at $Q^2=1$--$1.6$ GeV$^2$) to $\sigma_{\textrm{L}}$ are  about
from $-10\%$ to $-30\%$, to $\sigma_{\textrm{T}}$ are about $20\%$, and to $\sigma_{\textrm{LT,TT}}$ are much larger. By these analysis, we conclude that the TPE contributions in $ep\rightarrow en\pi^+$ at small $-t$ are important to extract the separated cross sections $\sigma_{\textrm{L,T,LT,TT}}$ and the electromagnetic magnetic form factor of $\pi^+$ in the experimental analysis.
\end{abstract}

\maketitle

\section{Introduction}

In the last two decades, the two-photon-exchange (TPE) effects in $ep\rightarrow ep$ have attracted much interest due to their importance in the extraction of  the electromagnetic (EM) form factors of protons. Many model-dependent methods have been used to estimate the TPE contributions in $ep\rightarrow ep$ such as the hadronic model \cite{hadronic model}, GPD method \cite{GPD method}, perturbative QCD (pQCD) calculation \cite{pQCD method}, dispersion-relation approach \cite{dispersion relation-1,dispersion relation-2}, SCEF method \cite{SCEF}, and phenomenological parametrization \cite{phenomenological parametrizations}.  Among all these methods, the dispersion-relation approach for $ep$ scattering gives the most reliable results in the region with medium momentum transfer, and the cost is that it needs to  continue the physical quantity analytically into the unphysical region and take some experimental data as inputs to fix the subtraction constant \cite{dispersion relation-2}. Furthermore, the difference between the dispersion-relation approach and the hadronic model can be expressed as a polynomial function on the squared center-of-mass energy. In some special cases, the two methods give the same results.

Due to the important contributions of the TPE corrections in $ep\rightarrow ep$, similar TPE corrections in $e^+e^-\rightarrow p\overline{p}$ \cite{TPE-ee-ppbar}, $e\pi\rightarrow e\pi$ \cite{TPE-epi-epi}, $ep\rightarrow eN\pi$\cite{Afanasev2013}, $\mu p \rightarrow \mu p$ \cite{TPE-mup-mup}, and $ep \rightarrow e\Delta \rightarrow ep\pi^0$ \cite{TPE-ep-eppi0} are studied, aiming at the precise extraction from the experimental data of  the EM form factor of protons in the timelike region, the EM form factor of pions in the spacelike region, and the EM transition form factors of $\gamma^*N\Delta$ in the spacelike region.

Experimentally, the extraction of the EM form factor of pion via $e\pi\rightarrow e\pi$ is limited at very small $Q^2$ with $Q^2\equiv -q^2$  and with $q$ being the four-momentum transfer because there is no free pion target.  The electromagnetic production of pions in $ep\rightarrow en\pi^+$ is usually used to extract the EM form factor of pions \cite{electro-pion-production-Cornell,electro-pion-production-DESY,electro-pion-production-JLab-1,electro-pion-production-JLab-2,electro-pion-production-JLab-3}. It is a natural question that how large  the TPE contributions in this process and how large their corrections to the extracted EM form factor of pion are. In this work, we estimate the TPE contributions in this process within the hadronic model and analyze the TPE corrections to the separated cross sections which are used to determine the EM form factor of pions.

We organize the paper as follows: In Sec.  II we describe the basic formulas of our calculation under the pion-dominance approximation, in Sec. III we express the physical amplitude as a sum of two invariant amplitudes and discuss the infrared (IR) property of the TPE amplitude, in Sec. IV we express the unpolarized differential cross section by the coefficients of the invariant amplitudes, in Sec. V we present the numerical results for the TPE corrections to the amplitude, to the unpolarized differential cross section and to the separated cross sections $\sigma_{\textrm{L,T,LT,TT}}$. A detailed discussion on these numerical results and the conclusion from these numerical results are also given.

\section{Basic Formula for $ep\rightarrow en\pi^+$}

\begin{figure}[htbp]
\includegraphics[width=8cm]{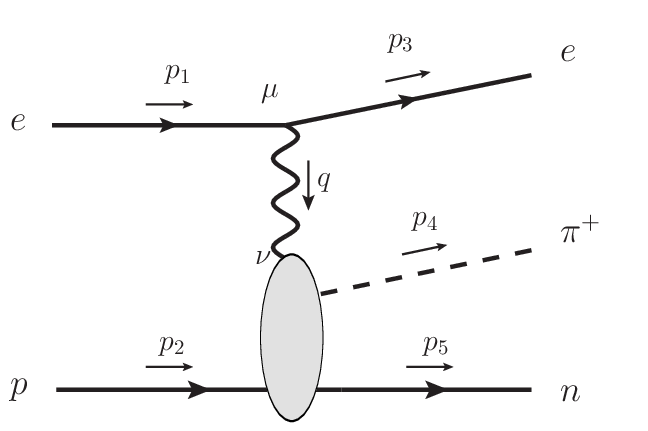}
\caption{$ep\rightarrow en\pi^+$ under the one-photon exchange.}
\label{figure:momenta}
\end{figure}

Under the one-photon exchange (OPE) approximation, the $ep\rightarrow en\pi^+$ process can be separated into two subprocesses, $e\rightarrow e\gamma^*$ and $\gamma^*p\rightarrow n\pi^+$, shown in Fig. \ref{figure:momenta} where we label the momenta of initial electron, initial proton, final electron, final pion, and final neutron as $p_{1}-p_{5}$, respectively, and for simplicity we define the following five independent Lorentz-invariant variables: $s \equiv (p_1+p_2)^2$, $Q^2\equiv-q^2\equiv -(p_1-p_3)^2$, $W\equiv[(p_4 + p_5)^2]^{1/2}$, $t\equiv (p_2-p_5)^2$, and $p_{14}\equiv p_1\cdot p_4$.

\begin{figure}[htbp]
\includegraphics[width=12cm]{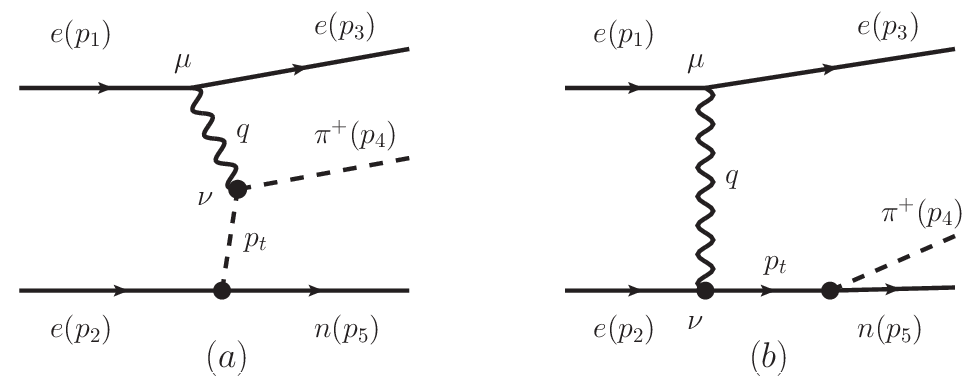}
\caption{Diagrams for $ep\rightarrow en\pi^+$  under the one-photon exchange with (a) the pion exchange diagram and (b) the elastic $s$-channel diagram.}
\label{figure:OPE-diagrams}
\end{figure}

The dynamics of the subprocess $e\rightarrow e\gamma^*$ is clear while the dynamics of the subprocess $\gamma^*p\rightarrow n\pi^+$ is very complex. In this work we limit our discussion on the momenta region with $Q^2$ small, $-t \approx 0$ and $W$ far away from the resonances. In this region, one can estimate the subprocess $\gamma^*p\rightarrow n\pi^+$ in the hadronic level as an approximation and can expect that the $\pi$ exchange diagram showed in Fig. \ref{figure:OPE-diagrams}(a) may give the most important contribution due to the large enhancement from the pion propagator.
In Fig. \ref{figure:OPE-diagrams}, the $s$-channel diagram is also presented to keep the gauge invariance. The sum of the $t$-channel and $s$-channel is gauge invariant. The contribution from the $u$-channel diagram with the neutron as the intermediate state is also gauge invariant independently and is small at low $-t$ since the electric charge of neutron is zero and it couples to photon via $F_{\mu\nu}$. We neglect this contribution in the following and only consider the sum of the $t$-channel and $s$-channel.

The unpolarized differential cross section at small $-t$ is usually used to determine the EM form factor of pions. Different from $e\pi^+\rightarrow e\pi^+$ process where the EM form factor of pions can be extracted from the total cross section directly,
the EM form factor can not be extracted directly from the total unpolarized cross section of $ep\rightarrow en\pi^+$ and should be extracted via the $\phi_{\pi}$-dependence of the unpolarized differential cross section. The TPE contributions may change the angle dependence of the unpolarized
differential cross section and then effect the extraction of the EM form factor in an indirect and nontrivial way.

\begin{figure}[htbp]
\includegraphics[width=15cm]{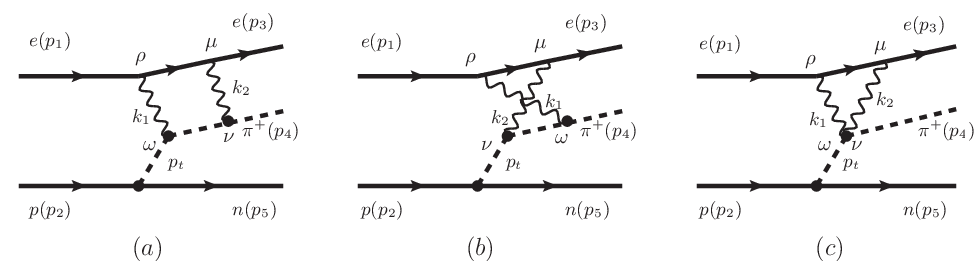}
\includegraphics[width=15cm]{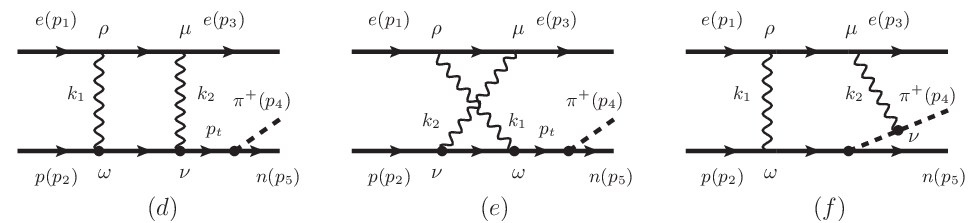}
\caption{Diagrams for $ep\rightarrow en\pi^+$  with two-photon exchange where  panels (a)-(c) correspond to the $\pi$ exchange $t$-channel one-photon exchange diagram and panles (d)-(f) correspond to the $s$-channel one-photon exchange diagram.}
\label{figure:TPE-diagrmas}
\end{figure}

When we discuss the TPE effects, the contributions from the corresponding TPE diagrams showed in  Fig. \ref{figure:TPE-diagrmas} should be considered where the TPE contributions from Fig. \ref{figure:TPE-diagrmas}(d)-3(f) are also included to keep the gauge invariance. In principle, when $Q^2\approx 1$ GeV$^2$ the contributions from the elastic state and the inelastic states such as $2\pi(\rho,\sigma),3\pi$  between the two photons may both give the vital contributions. In this work, as a first step we limit our discussion on the contributions from the elastic state since naively the transition form factors $\gamma^*\pi\rho,\gamma^*\pi\sigma $ are much smaller than the EM form factor $\gamma^*\pi\pi$ when $Q^2$ increase.

When taking Feynman gauge and limiting the discussion on the small $-t$, the contributions from the diagrams Figs. \ref{figure:OPE-diagrams}(a) and  \ref{figure:TPE-diagrmas}(a)-3(c) are the most important in the OPE and TPE levels, respectively. Since we are only interested in the property of the TPE corrections or the ratio of the TPE contributions to the OPE contributions, in the following discussion we only consider the contributions from Figs. \ref{figure:OPE-diagrams}(a) and \ref{figure:TPE-diagrmas}(a)-(c). Such simplification has an advantage that the TPE contributions have a very simple form in the amplitude level.

Under the above approximation, the ratio of the TPE contributions to the OPE contributions is not dependent on the interactions between pions, protons, and neutrons since the relative TPE corrections are only dependent on the subprocess $e\pi^*\rightarrow e\pi$. In the practical calculation we simply take the interaction between pions, protons, and neutrons as iso-scalar type. We use the interactions constructed in Ref. \cite{zhouhq2011-pion-photon-interaction} to describe the interactions between
the pion and the photon.

Taking the Feynman gauge, one has
\begin{eqnarray}
\mathcal{M}_{1\gamma}^{(a)} &=& - i  \bar{u}_e(p_3)(-ie\gamma^{\mu}) u_e(p_1)\  \bar{u}_n(p_5)(-g_0\gamma_5) u_p(p_2)\Gamma^{\nu}(p_4,p_t) S_{\pi}(p_t)D_{\mu \nu}(p_1-p_3),\nonumber \\
\mathcal{M}_{2\gamma}^{(a)} &=& - i \int \frac{d^4 k_1}{(2\pi)^4}\bar{u}_e(p_3)(-ie\gamma^{\mu})S_F(p_1-k_1)(-ie\gamma^{\rho}) u_e(p_1)\ \bar{u}_n(p_5)(-g_0\gamma_5) u_p(p_2) \Gamma^{\nu}(p_4,p_4-k_2)  \nonumber \\
&&~~~~~~ S_{\pi}(p_4-k_2)\Gamma^{\omega}(p_4-k_2,p_t)S_{\pi}(p_t) D_{\mu \nu}(k_2)D_{\rho \omega}(k_1), \nonumber \\
\mathcal{M}_{2\gamma}^{(b)} &=& - i \int \frac{d^4 k_1}{(2\pi)^4}\bar{u}_e(p_3)(-ie\gamma^{\mu})S_F(p_1-k_1)(-ie\gamma^{\rho}) u_e(p_1)\ \bar{u}_n(p_5)(-g_0\gamma_5) u_p(p_2) \Gamma^{\omega}(p_4,p_4-k_1) \nonumber \\
&&~~~~~~ S_{\pi}(p_4-k_1)\Gamma^{\nu}(p_4-k_1,p_t)S_{\pi}(p_t) D_{\mu \nu}(k_2)D_{\rho \omega}(k_1),  \nonumber \\
\mathcal{M}_{2\gamma}^{(c)} &=& - i \int \frac{d^4 k_1}{(2\pi)^4}\bar{u}_e(p_3)(-ie\gamma^{\mu})S_F(p_1-k_1)(-ie\gamma^{\rho}) u_e(p_1)\ \bar{u}_n(p_5)(-g_0\gamma_5) u_p(p_2)\Lambda^{\omega\nu}(k_1,k_2)S_{\pi}(p_t) \nonumber\\
&&~~~~~~   D_{\mu \nu}(k_2)D_{\rho \omega}(k_1),
\label{eq:amplitudes-OPE-TPE}
\end{eqnarray}
with
\begin{eqnarray}
S_F(k) &=& \frac{i(\sla{k}+m_e)}{k^2-m_e^2+i\epsilon}, \nonumber \\
S_{\pi}(k) &=& \frac{i}{k^2-m_\pi^2+i\epsilon}, \nonumber \\
D_{\mu\rho}(k) &=& \frac{-i}{k^2+i\epsilon}g_{\mu\rho},
\end{eqnarray}
and
\begin{eqnarray}
\Gamma^{\mu}(p_f,p_i) &=& ie\left[(1+f(k^2)k^2)(p_f+p_i)^{\mu}-f(k^2)(p_f^2-p_i^2)k^{\mu} \right] \nonumber\\
\Lambda^{\mu\nu}(k_1,k_2)&=& 2ie^2\left[ g^{\mu\nu}+f(k_1^2)(k_1^2 g^{\mu\nu}-k_1^{\mu}k_1^{\nu}) + f(k_2^2)(k_2^2 g^{\mu\nu}-k_2^{\mu}k_2^{\nu}) \right],
\end{eqnarray}
where $e=-|e|$, $k\equiv p_f-p_i$, and $f(k^2)$ describes the EM form factor of pion $F_\pi(k^2)$ and has the relation
\begin{eqnarray}
F_\pi(k^2)&=&1+k^2f(k^2).
\label{relation}
\end{eqnarray}

In Ref. \cite{dispersion relation-2}, the authors discussed and compared the TPE contributions calculated from the hadronic model and the dispersion-relation in detail in the $ep$ case. Their discussions hint that the two methods give the same results when the interaction between the point-like particles is traditionally renormalized  and give different results when the interaction between the point-like particles is traditionally nonrenormalized, respectively. In $ep$ scattering, the pure electric interaction belongs to the former, and the pure magnetic interaction belong to the latter. This property is natural since a new contact interaction should be introduced when the interaction leads to a traditional nonrenormalized UV divergence. Such contact interaction includes an undetermined finite contribution in polynomial form and should be also considered in the hadronic model. This reason can also explain the behavior of the difference between the dispersion-relation approach and the hadronic model in the $ep$ case. In the dispersion-relation approach, such finite contribution is fixed by the experimental data or by the asymptotic behavior in high energy predicted by other methods. In the $e\pi$-interaction case, the situation is different from the $ep$ case since now the interaction for a point-like charged pseudoscalar particle is renormalized. On the other hand the gauge invariance results in a contact term. Such a contact term gives a pure real contribution and does not appear in the $ep$ case. These two properties prompts us to use the dynamical hadronic model to calculate the TPE contribution in $ep\rightarrow en\pi^+$ when only the elastic state is included.

\section{The IR divergence of the amplitude}

Generally, the amplitudes given in Eq. (\ref{eq:amplitudes-OPE-TPE}) can be expressed in the following simple form:
\begin{eqnarray}
\mathcal{M}_{1\gamma} &\equiv& \mathcal{M}_{1\gamma}^{(a)} = c_1^{(1\gamma)} \mathcal{M}_1 + c_2^{(1\gamma)} \mathcal{M}_2,\nonumber \\
\mathcal{M}_{2\gamma} &\equiv& \mathcal{M}_{2\gamma}^{(a+b+c)} = c_1^{(2\gamma)} \mathcal{M}_1 + c_2^{(2\gamma)} \mathcal{M}_2,
\label{eq:amplitude-definition}
\end{eqnarray}
with
\begin{eqnarray}
\mathcal{M}_1 &\equiv& i\bar{u}(p_3,m_e)(2\sla{p}_4+\sla{p}_3-\sla{p}_1)u(p_1,m_e) \  \bar{u}(p_5,m_n)\Gamma_5 u(p_2,m_p), \nonumber\\
\mathcal{M}_2 &\equiv& i\bar{u}(p_3,m_e)u(p_1,m_e) \  \bar{u}(p_5,m_n)\Gamma_5 u(p_2,m_p),
\label{eq:invariant-amplitude}
\end{eqnarray}
with $\Gamma_5\equiv g_{\pi NN}\gamma_5$ being the vertex of the $\pi NN$ isoscalar interaction. The coefficients $c^{(1\gamma)}_1$ and $c^{(1\gamma)}_2$ can be easily gotten and are expressed as
\begin{eqnarray}
c_1^{(1\gamma)} &=& \frac{4\pi  \alpha_e F_{\pi}(q^2)}{Q^2(t-m_{\pi}^2)}, \nonumber\\
c_2^{(1\gamma)} &=& 0,
\end{eqnarray}
with $\alpha_e\equiv e^2/4\pi$.

When taking the approximation $m_e=0$ one has $c_{2}^{(2\gamma)}=0$  due to the symmetry and our numerical results also show this property. The expressions for $c_1^{(2\gamma)}$ and $c_2^{(2\gamma)}$ are complex; even the form factor $f(k^2)$ is taken as a simple monopole form. A general property is that there is only IR divergence  in $c_{1}^{(2\gamma)}$. The detailed analysis shows that the IR divergence comes from the diagrams in Figs. 3(a) and 3(b), and the corresponding IR divergence \cite{Pacakge X} in $c_1^{(2\gamma)}$ by keeping $m_e$ in the propagator
can be expressed as
\begin{eqnarray}
c_{1,\text{IR}}^{(2\gamma, a)} &=& \frac{-2\alpha_e^2F_{\pi}(q^2)  }{Q^2(t-m_{\pi}^2)}  \frac{ a\log\big[\frac{-a+\sqrt{a^2-4m_e^2 m_{\pi}^2}}{2m_e m_{\pi}}\big]}{\sqrt{a^2-4m_e^2 m_{\pi}^2}} \log\frac{\lambda^2}{\mu_c^2} \nonumber\\
&\approx& \frac{-2\alpha_e^2 F_{\pi}(q^2)}{Q^2(t-m_{\pi}^2)} \big(\log\frac{m_em_\pi}{a}+i\pi\big)\log\frac{\lambda^2}{\mu_c^2},
\end{eqnarray}
and
\begin{eqnarray}
c_{1,\text{IR}}^{(2\gamma, b)} &=& \frac{-2\alpha_e^2F_{\pi}(q^2)}{Q^2(t-m_{\pi}^2)}  \frac{ p_{14}\log\big[\frac{p_{14}+\sqrt{p_{14}^2-m_e^2 m_{\pi}^2}}{m_e m_{\pi}}\big]}{\sqrt{p_{14}^2-m_e^2 m_{\pi}^2}} \log\frac{\lambda^2}{\mu_c^2} \nonumber \\
&\approx& \frac{-2\alpha_e^2 F_{\pi}(q^2)}{Q^2(t-m_{\pi}^2)} \log\frac{2p_{14}}{m_em_\pi}\log\frac{\lambda^2}{\mu_c^2},
\end{eqnarray}
where $a\equiv 2p_{14}+Q^2+t-m_{\pi}^2$, $\lambda$ is the introduced infinitesimal mass of photon, $\mu_c$ is an energy scale to keep the result dimensionless, and the properties that $a>0,p_{14}>0$ in the physical region are used in the expansions on $m_e$. The results show that the full IR divergence is free from $m_e$. The above IR divergence should be included in any experimental data analysis when the real radiative corrections are included.

In $ep\rightarrow ep$ process, the contribution from the TPE diagrams under the soft momentum approximation which includes the IR divergence is usually estimated via the classical Mo-Tsai's soft approximation \cite{IR-Mo-and-Tsai} in the experimental analysis. In this approximation the soft TPE contribution is calculated by taking the momentum of one photon as zero both in the numerator and one of the denominators of the propagators. In Ref. \cite{IR-Maximon-and-Tjon}, Maximon and Tjon suggest another approximation to estimate the soft TPE contribution. In their estimation, the soft contribution is calculated by taking momentum of one photon as zero only in the numerator. The analytical expressions in the latter method can be get in $ep\rightarrow ep$ or $e\pi\rightarrow e\pi$. In the $ep\rightarrow en\pi^+$ process, the intermediate pion is off-shell, which introduces an additional variable $t$, the analytical expressions under the above soft approximation are very complex and we do not go to show them. To show the TPE corrections from the finite momentum transform, we define the IR-free TPE contributions as follows:
\begin{eqnarray}
c_{1,\text{fin}}^{(2\gamma)} \equiv c_1^{(2\gamma)}-(c_{1,\text{IR}}^{(2\gamma,a)}+c_{1,\text{IR}}^{(2\gamma,b)}),\nonumber\\
c_{1,\text{Tsai}}^{(2\gamma)} \equiv c_1^{(2\gamma)}-(c_{1,\text{Tsai}}^{(2\gamma,a)}+c_{1,\text{Tsai}}^{(2\gamma,b)}),\nonumber\\
c_{1,\text{Tjon}}^{(2\gamma)} \equiv c_1^{(2\gamma)}-(c_{1,\text{Tjon}}^{(2\gamma,a)}+c_{1,\text{Tjon}}^{(2\gamma,b)}),
\end{eqnarray}
where the indexes $\textrm{Tsai}$ and $\textrm{Tjon}$ refer to the corresponding contributions by the Mo-Tsai's method and Maximon-Tjon's method, respectively. In the practical calculation with the experimental momenta as inputs, we find that the results $c_{1,\text{fin}}^{(2\gamma)}$ with $\mu_c=1$ GeV are close to $c_{1,\text{Tsai}}^{(2\gamma)}$, while they are much different from $c_{1,\text{Tjon}}^{(2\gamma)}$. In the following discussion we use $c_{1,\text{Tsai}}^{(2\gamma)}$ to show the TPE contributions, since the soft contributions by Mo-Tsai's method are usually used to analyze the data sets in experiments.

\section{The unpolarized cross section}

Using the general expression of the amplitudes (\ref{eq:amplitude-definition}) and  (\ref{eq:invariant-amplitude}), one can get the  expressions of the unpolarized differential scattering cross sections as follows:
\begin{eqnarray}
\frac{d^5 \sigma^{1\gamma}_{un}}{dE_{e'} d\Omega_{e'} d\Omega_{\pi}} &\propto & \sum_{spin} \mathcal{M}_{1\gamma}\mathcal{M}_{1\gamma}^* \nonumber\\
&=& 8 |c_1^{(1\gamma)}|^2 (-t) \left[ 8p_{14}^2+4(Q^2+ t - m_{\pi}^2)p_{14}-2m_{\pi}^2 Q^2 \right],
\end{eqnarray}
\begin{eqnarray}
\frac{d^5 \sigma^{2\gamma}_{un}}{dE_{e'} d\Omega_{e'} d\Omega_{\pi}} &\propto& \sum_{spin} 2\textrm{Re}[\mathcal{M}_{2\gamma}\mathcal{M}_{1\gamma}^*] \nonumber\\
&=&2  {\textrm{Re}}\Big\{8c_1^{(1\gamma)}c_{1,\text{fin}}^{(2\gamma)} (-t) \left[ 8p_{14}^2+4(Q^2+ t - m_{\pi}^2)p_{14}-2m_{\pi}^2 Q^2\right] \nonumber\\
&& + 8 m_e c_1^{(1\gamma)}c_2^{(2\gamma)} (-t) (-m_{\pi}^2+4p_{14}+Q^2+t)\Big\},
\label{equ:Ana TPE cross section}
\end{eqnarray}
where $E_{e'}$ is the energy of final electron in the laboratory frame, $\Omega_{e'}$ is the angle of the final electron in the laboratory frame, $\Omega_\pi$ is the angle of the pion in the center frame of the pion and final proton, and we have taken $c_1^{(1\gamma)}$ as real. From Eq. (\ref{equ:Ana TPE cross section}) one can also see that the contribution from $c_2^{(2\gamma)}$ can be neglected when making the approximation $m_e=0$.

The unpolarized cross sections above can be written as
\begin{eqnarray}
\frac{d^5 \sigma^X_{un}}{dE_{e'} d\Omega_{e'} d\Omega_{\pi}} \equiv \Gamma_{\nu} J(t,\phi_{\pi} \rightarrow \Omega_{\pi}) \frac{d^2 \sigma^X_{un}}{dt d\phi_{\pi}},
\end{eqnarray}
where $X$ refers to $1\gamma$ or $2\gamma$, $J(t,\phi_{\pi} \rightarrow \Omega_{\pi})= \frac{dt}{\sin\theta_{\pi} d\theta_{\pi}}$,  and $\Gamma_{\nu} = \frac{\alpha_{e}}{2\pi^2} \frac{E_{e'}}{E_e} \frac{W^2-m_p^2}{2m_pQ^2} \frac{1}{1-\epsilon}$ is the virtual photon flux factor with  $E_e$ being the energy of initial electron in the laboratory frame, $m_p$ the mass of the proton, and $\epsilon$  the longitudinal polarization of the virtual photon, whose definition can be found in the appendix.  According to the dependence on $\phi_{\pi}$ and $\epsilon$, the OPE cross section $d^2 \sigma_{un}^{1\gamma}/dt d\phi_{\pi}$ can be separated into four terms as follows:
\begin{eqnarray}
2\pi \frac{d^2\sigma^{1\gamma}_{un}}{dt d\phi_{\pi}} &=& \epsilon \frac{d \sigma_{\textrm{L}}^{1\gamma}}{dt} + \frac{d \sigma_{\textrm{T}}^{1\gamma}}{dt} + \sqrt{2\epsilon (\epsilon+1)}\frac{d \sigma_{\textrm{LT}}^{1\gamma}}{dt} \cos\phi_{\pi}+\epsilon\frac{d \sigma_{\textrm{TT}}^{1\gamma}}{dt} \cos2\phi_{\pi}\nonumber\\
&\equiv& \epsilon \sigma_{\textrm{L}}^{1\gamma} +  \sigma_{\textrm{T}}^{1\gamma} + \sqrt{2\epsilon (\epsilon+1)} \sigma_{\textrm{LT}}^{1\gamma} \cos\phi_{\pi}+\epsilon \sigma_{\textrm{TT}}^{1\gamma} \cos2\phi_{\pi},
\label{eq:OPE-cs-separared-form}
\end{eqnarray}
where the four separated cross sections $d \sigma_{\textrm{L,T,LT,TT}}^{1\gamma}/dt$ shortly written as $\sigma_{\textrm{L,T,LT,TT}}^{1\gamma}$ are only depend on $Q^2$, $W$, and $\theta_{\pi}$.

When one takes $m_e=0$ in Eq. (\ref{equ:Ana TPE cross section}), one can see that the TPE cross section $d^2 \sigma_{un}^{2\gamma}/dt d\phi_{\pi}$ has the same form with OPE cross section. After using the variables $Q^2,W,\epsilon,\theta_\pi$, and $\phi_\pi$ to express the cross section one can see that $d^2 \sigma_{un}^{2\gamma}/dt d\phi_{\pi}$  has the same $\phi_\pi$ dependence with $d^2 \sigma_{un}^{1\gamma}/dt d\phi_{\pi}$  and can also be separated into the same form as Eq. (\ref{eq:OPE-cs-separared-form}) but now the four corresponding separated cross sections $\sigma_{\textrm{L,T,LT,TT}}^{2\gamma}$ are dependent on $Q^2$, $W$, $\theta_{\pi}$, and $\epsilon$.

\section{The numerical results and discussion}
In the practical calculation, we take the input form factor $F_{\pi}(q^2)$ as the monopole from which is used in \cite{Blunden2010-pion-form-factor}  and  \cite{zhouhq2011-pion-photon-interaction}.
\begin{eqnarray}
F_\pi(q^2)=\frac{-\Lambda^2}{q^2-\Lambda^2},
\end{eqnarray}
with $\Lambda=0.77$ GeV.  We use the packages FEYNCALC \cite{FenyCalc} and LOOPTOOLS \cite{LoopTools} to carry out the analytical
and numerical calculations, respectively. For comparison, we take the experiment kinematics in JLab $F_\pi$ \cite{electro-pion-production-JLab-2} with $Q^2= 1 \text{ GeV}^2$ and $Q^2=1.6 \text{ GeV}^2$ at $W=1.95 \text{ GeV}$ as examples to show the TPE contributions.

\subsection{Two-photon-exchange contributions to the amplitude $c_{1,\text{Tsai}}^{(2\gamma)}/c_1^{(1\gamma)}$}
\begin{figure*}[htbp]
\centering
\includegraphics[width=15cm]{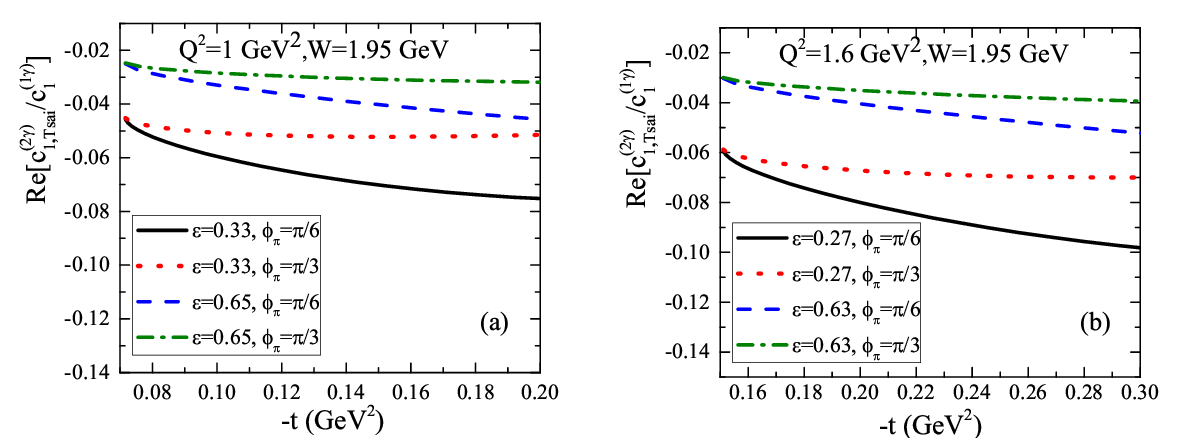}
\caption{Numeric results for $\textrm{Re}[c_{1,\text{Tsai}}^{(2\gamma)}/c_1^{(1\gamma)}]$ vs $-t$ at fixed $Q^2,W,\epsilon$, and $\phi_\pi$. The left panel is the result with $Q^2=1$ GeV$^2$ and the right panel is the result with $Q^2=1.6$ GeV$^2$.}
\label{figure:TPE-to-c2-on-t-Re}
\end{figure*}

The $-t$ dependence of the TPE correction $\textrm{Re} [c_{1,\text{Tsai}}^{(2\gamma)}/c_1^{(1\gamma)}]$ is presented in Fig. \ref{figure:TPE-to-c2-on-t-Re} where the left and right panels are corresponding to $Q^2=1 \text{ GeV}^2$ and $Q^2=1.6 \text{ GeV}^2$, respectively. The (blue) dashed curves and the (olive) dash-dotted  curves refer to the results at $\phi_{\pi}=\pi/6$  and $\phi_{\pi}=\pi/3$ with $\epsilon=0.65$ or $0.63$, the (black) solid curves and the (red) dotted curves are associated with $\epsilon=0.33$ or $0.27$. The results clearly show that the absolute magnitude of TPE corrections $\textrm{Re}[c_{1,\text{Tsai}}^{(2\gamma)}/c_1^{(1\gamma)}]$ at $\phi_{\pi} = \pi/6$ increase when $-t$ increases while the corrections at $\phi_{\pi}=\pi/3$ are not sensitive to $-t$. Another interesting property is that the TPE corrections at very small $-t$ are not sensitive to $\phi_\pi$  while the TPE corrections at large $-t$ are sensitive to $\phi_\pi$.

At $\phi_{\pi}=\pi/6$, one can see that the TPE corrections at small $\epsilon$ range about from $-4\%$ to $-6\%$ at small $-t$ and reach about from $-7\%$ to $-10\%$ at large $-t$ at $Q^2=1.0$ and $1.6$ GeV$^2$, respectively. The magnitude at small $-t$  and small $\epsilon$ is similar to the TPE corrections in $e\pi\rightarrow e\pi$. These properties suggest that the $-t$ dependence of the TPE corrections at small $\phi_{\pi}$ is relatively important.

\begin{figure*}[htbp]
\centering
\includegraphics[width=15cm]{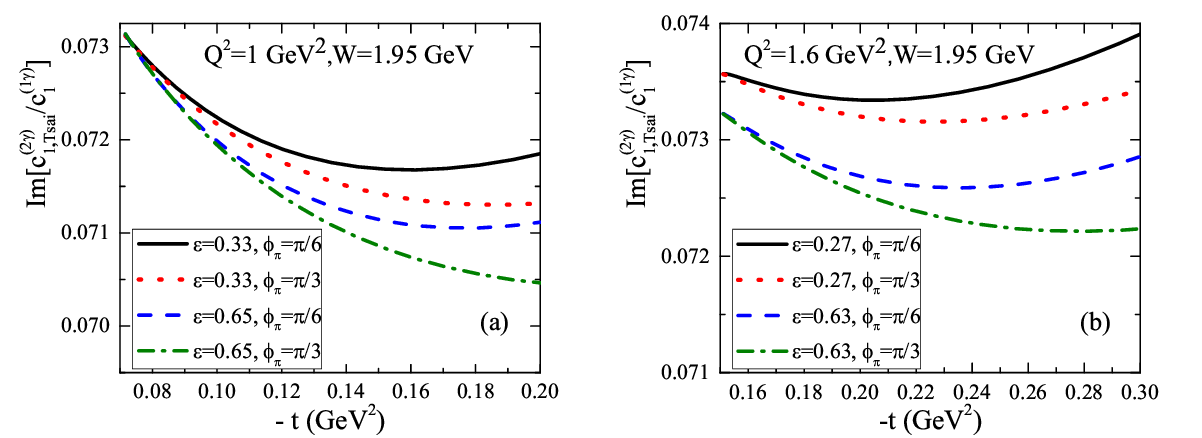}
\caption{Numeric results for $\textrm{Im}[c_{1,\text{Tsai}}^{(2\gamma)}/c_1^{(1\gamma)}]$ vs $-t$ at fixed $Q^2,W,\epsilon$, and $\phi_\pi$. The left panel is the result with $Q^2=1$ GeV$^2$ and the right panel is the result with $Q^2=1.6$ GeV$^2$.}
\label{figure:TPE-to-c2-on-t-Im}
\end{figure*}

The $-t$ dependence of the imaginary parts of the TPE corrections, $\textrm{Im} [c_{1,\text{Tsai}}^{(2\gamma)}/c_1^{(1\gamma)}]$,  is presented in Fig. \ref{figure:TPE-to-c2-on-t-Im} where the same definitions as in Fig. \ref{figure:TPE-to-c2-on-t-Re} are used for the curves. The results show an interesting and important property: the TPE corrections to the imaginary part are not sensitive to $Q^2,\phi_\pi,-t$, and $\epsilon$ at $W=1.95$ GeV and are almost about 7\%.


\begin{figure*}[htbp]
\centering
\includegraphics[width=15cm]{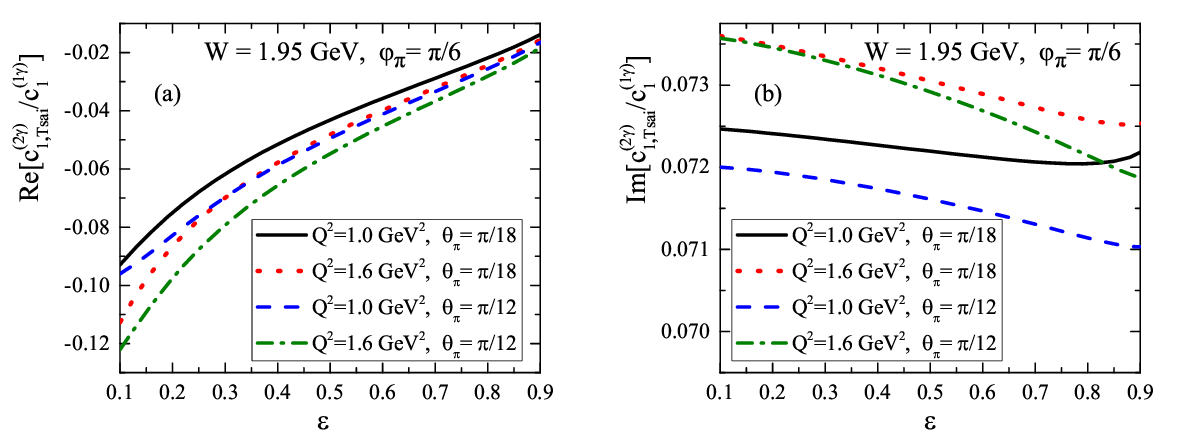}
\caption{Numeric results for $c_{1,\text{Tsai}}^{(2\gamma)}/c_1^{(1\gamma)}$ vs $\epsilon$ at fixed $Q^2,W, \theta_{\pi}$, and $\phi_\pi$. The left panel is the result for the real part and the right panel  is the result for the imaginary part.}
\label{figure:TPE-to-c2-on-epsilon}
\end{figure*}

The $\epsilon$ dependence of the TPE corrections $c_{1,\text{Tsai}}^{(2\gamma)}/c_1^{(1\gamma)}$ is presented in Fig. \ref{figure:TPE-to-c2-on-epsilon} where $\theta_{\pi} $ is taken as $\pi/18,\pi/12$ and $-t$ is limited within the experimental data sets. The (black) solid curves and the (red) dotted curves refer to the results  with $\theta_{\pi} =\pi/18$ at $Q^2=1$ and $1.6 \text{ GeV}^2$, respectively. The (blue) dashed curves and the (olive) dash-dotted curves are associated with $\theta_{\pi} =\pi/12$. The results clearly show that the absolute magnitude of $\textrm{Re}[c_{1,\text{Tsai}}^{(2\gamma)}/c_1^{(1\gamma)}]$ decreases when $\epsilon$ increases. This is a general property of the TPE corrections. At $\epsilon=0.1$ the TPE corrections $\textrm{Re}[c_{1,\text{Tsai}}^{(2\gamma)}/c_1^{(1\gamma)}]$ reach about $-9\%$ and $-12\%$  at $Q^2=1$GeV$^2$ and 1.6 GeV$^2$, respectively.  The results in the right panel clearly show that the TPE corrections to the imaginary part are not sensitive to $\epsilon$.

\subsection{Comparison between $c_{1,\text{fin}}^{(2\gamma)}/c_1^{(1\gamma)}$, $c_{1,\text{Tsai}}^{(2\gamma)}/c_1^{(1\gamma)}$ and $c_{1,\text{Tjon}}^{(2\gamma)}/c_1^{(1\gamma)}$}

\begin{figure*}[htbp]
\centering
\includegraphics[width=15cm]{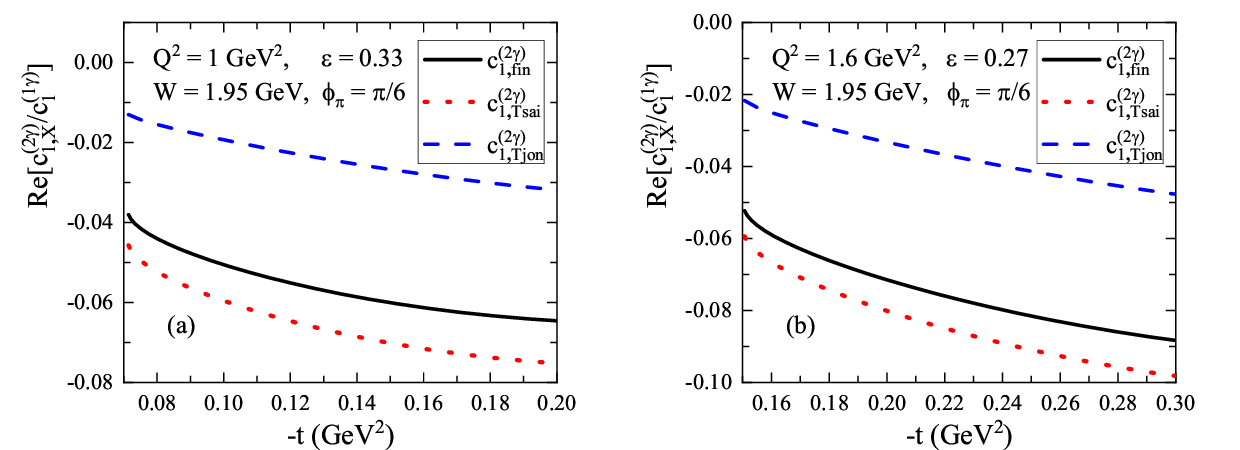}
\caption{Comparison between $\textrm{Re}[c_{1,\text{X}}^{(2\gamma)}]/c_1^{(1\gamma)}$ vs $-t$ at fixed $Q^2,W, \epsilon$, and $\phi_\pi$ where the index $\textrm{X}$ refers to fin, Tsai, and Tjon, respectively.}
\label{figure:Re-t-comparison}
\end{figure*}

\begin{figure*}[htbp]
\centering
\includegraphics[width=15cm]{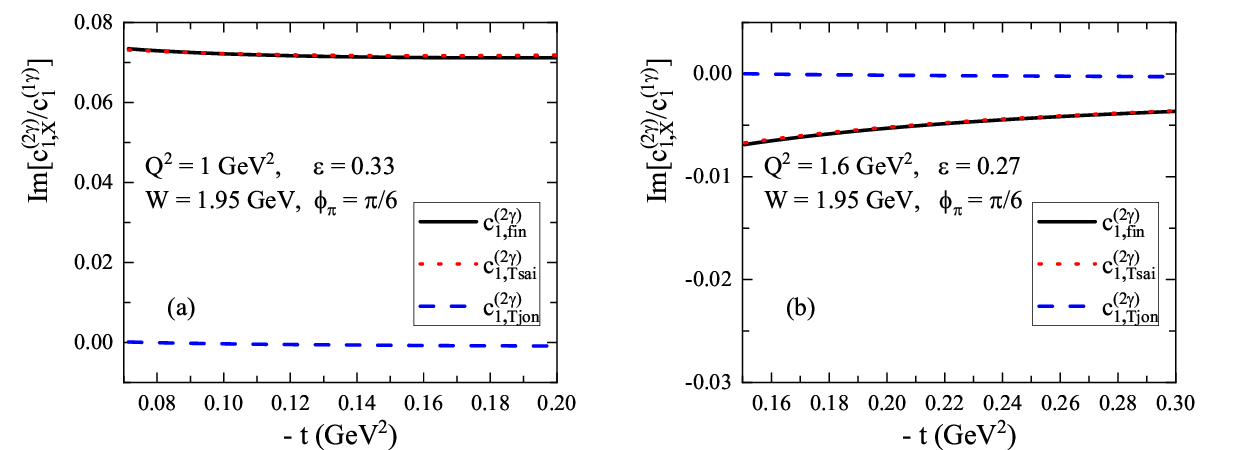}
\caption{Comparison between $\textrm{Im}[c_{1,\text{X}}^{(2\gamma)}]/c_1^{(1\gamma)}$ vs $-t$ at fixed $Q^2,W, \epsilon$, and $\phi_\pi$ where the index $\textrm{X}$ refers to fin, Tsai, and Tjon, respectively.}
\label{figure:Im-t-comparison}
\end{figure*}

\begin{figure*}[htbp]
\centering
\includegraphics[width=15cm]{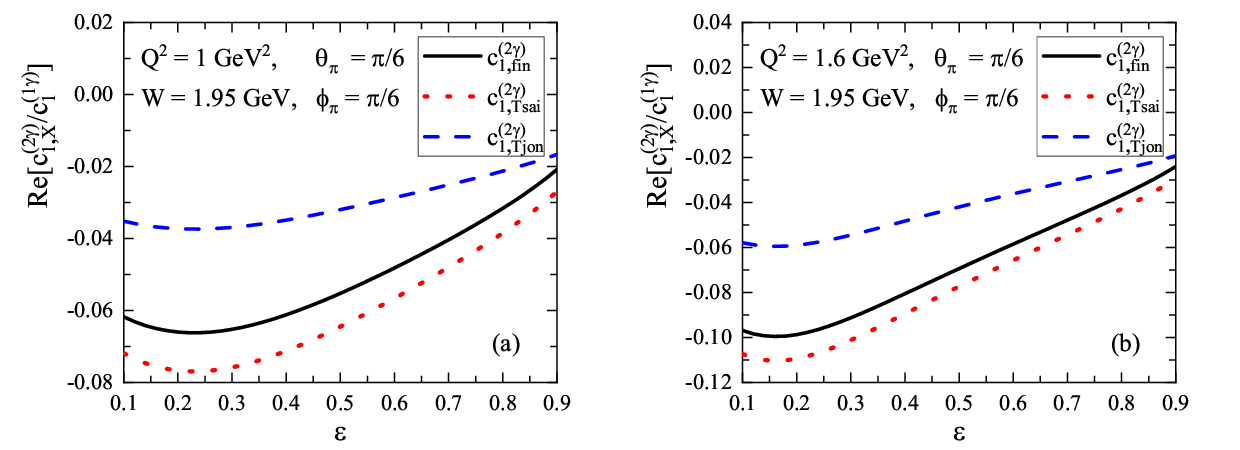}
\caption{Comparison between $\textrm{Re}[c_{1,\text{X}}^{(2\gamma)}]/c_1^{(1\gamma)}$ vs $\epsilon$ at fixed $Q^2,W, \theta_{\pi}$, and $\phi_\pi$ where the index $\textrm{X}$ refers to fin, Tsai, and Tjon, respectively.}
\label{figure:Re-epsilon-comparison}
\end{figure*}

\begin{figure*}[htbp]
\centering
\includegraphics[width=15cm]{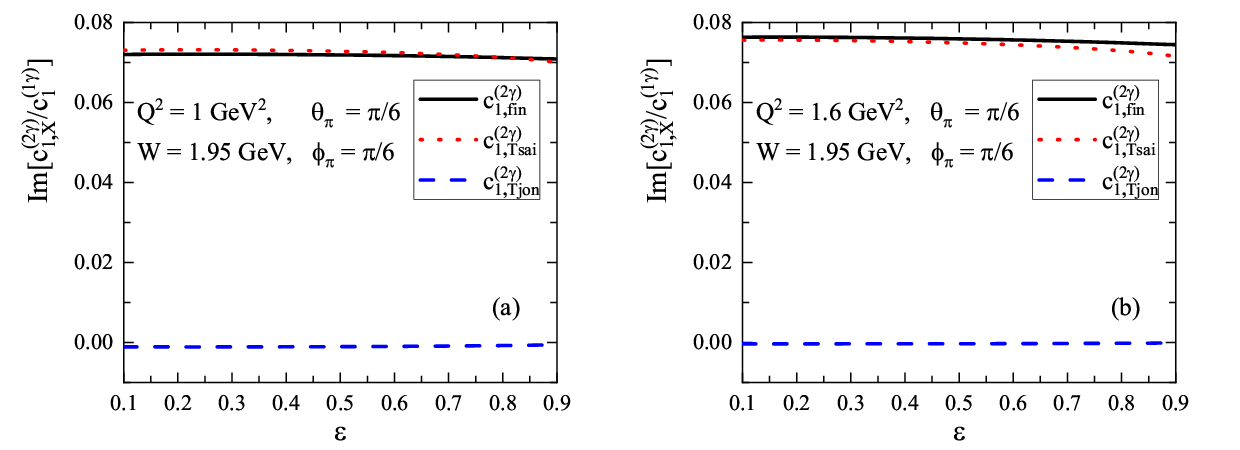}
\caption{Comparison between $\textrm{Im}[c_{1,\text{X}}^{(2\gamma)}]/c_1^{(1\gamma)}$ vs $\epsilon$ at fixed $Q^2,W, \theta_{\pi}$, and $\phi_\pi$ where the index $\textrm{X}$ refers to fin, Tsai, and Tjon, respectively.}
\label{figure:Im-epsilon-comparison}
\end{figure*}

In this section, we compare the results $c_{1,\text{X}}^{(2\gamma)}/c_1^{(1\gamma)}$ where the subindex $\textrm{X}$ refers to fin, Tsai, and Tjon, respectively. In Figs. \ref{figure:Re-t-comparison} and \ref{figure:Im-t-comparison} we present the results for  $c_{1,\text{X}}^{(2\gamma)}/c_1^{(1\gamma)}$ vs $-t$ at fixed $Q^2, W, \epsilon$ and $\phi_\pi$. In Figs. \ref{figure:Re-epsilon-comparison} and \ref{figure:Im-epsilon-comparison} we present the results for  $c_{1,\text{X}}^{(2\gamma)}/c_1^{(1\gamma)}$ vs $\epsilon$ at fixed $Q^2, W, \theta_{\pi}$ and $\phi_{\pi}$. These results clearly show that $c_{1,\text{fin}}^{(2\gamma)}$ are close to $c_{1,\text{Tsai}}^{(2\gamma)}$, but different from $c_{1,\text{Tjon}}^{(2\gamma)}$. Experimentally, the soft contributions by Mo-Tsai's method are usually used to analysis the data sets. Thus, we use  $c_{1,\text{Tsai}}^{(2\gamma)}/c_1^{(1\gamma)}$ to analysis the experimental data sets in the following.

\subsection{Two-photon-exchange corrections to unpolarized differential cross section }

To show the TPE corrections to the unpolarized differential scattering cross section, we define
\begin{eqnarray}
\delta^{2\gamma}_{un} \equiv \frac{d \sigma^{2\gamma}_{un}}{dtd\phi_\pi}/ \frac{d \sigma^{1\gamma}_{un}}{dt d\phi_\pi} = \frac{2\ \text{Re}[c_1^{(1\gamma)}c_{1,\text{Tsai}}^{(2\gamma)}]}{|c_1^{(1\gamma)}|^2} =2\ \text{Re}[\frac{c_{1,\text{Tsai}}^{(2\gamma)}}{c_1^{(1\gamma)}}],
\label{eq:TPE-to-cs}
\end{eqnarray}
where we have used the property that $c_1^{(1\gamma)}$ is real.  Equation (\ref{eq:TPE-to-cs}) means that the TPE corrections to the unpolarized cross sections are just twice the real part of the TPE corrections to the coefficient $c_{1}$. After considering this factor of two, one can see that the TPE corrections to the unpolarized cross section at small $\epsilon$, small $\phi_\pi$, and $Q^2=1$ GeV$^2$ can reach about $-10\%$, which is not small. Furthermore, the TPE corrections are sensitive to $\epsilon$, $\phi_\pi$, and $-t$ or $\theta_{\pi}$ when $Q^2$ and $W$ are fixed. Generally one can expect that these two properties may result in nontrivial effects when extracting some physical quantities from the angle dependence of the differential cross section.

When comparing with the TPE corrections in $e^+e^-\rightarrow p\overline{p}$ \cite{TPE-ee-ppbar}, $e\pi\rightarrow e\pi$ \cite{TPE-epi-epi}, $\mu p \rightarrow \mu p$ \cite{TPE-mup-mup}, and $ep \rightarrow e\Delta \rightarrow ep\pi^0$ at $W=1.232$ GeV \cite{TPE-ep-eppi0}, we can see that the absolute magnitude of the TPE corrections in $ep\rightarrow en\pi^+$ are much larger. This property can be understood by the fact that the intermediate pion with four-momentum $p_t\equiv p_5-p_2$ is off-shell which is different from the other processes. Naively, if $p_t^2=t$ goes to $m_\pi^2$, the TPE corrections to the coefficients should be same as the TPE corrections in the physical process $e\pi\rightarrow e\pi$. From Fig. \ref{figure:TPE-to-c2-on-t-Re}, one can see that the absolute magnitude of TPE corrections $\textrm{Re} [c_{1,\text{Tsai}}^{(2\gamma)}/c_1^{(1\gamma)}]$ decreases when $t$ increases in the region $t\subseteq[-0.2,-0.08]$ GeV$^2$.

\subsection{Two-photon-exchange corrections to separated cross sections $\sigma_{\textrm{L}}, \sigma_{\textrm{T}},\sigma_{\textrm{LT}}$ and $\sigma_{\textrm{TT}}$}

Experimentally, the separated cross sections $  \sigma_{\textrm{L}} $, $ \sigma_{\textrm{T}} $, $  \sigma_{\textrm{LT}} $, and $ \sigma_{\textrm{\textrm{TT}}} $ are usually extracted from the original experimental data $d\sigma^{Ex}_{un}/dtd\phi_\pi$ via Eq. (\ref{eq:OPE-cs-separared-form}) and are then used to determine the EM form factor of $\pi^+$. Since the TPE corrections to the unpolarized cross section are not small and sensitive on the angles, one should be careful in the separation. In this section, we analyze the TPE corrections to the separated cross sections.

When considering the TPE contribution, one has
\begin{eqnarray}
\frac{d\sigma^{Ex}_{un}}{dtd\phi_\pi} =\frac{ d\sigma^{ph,1\gamma}_{un}}{dtd\phi_\pi}(1+ \delta^{ph,2\gamma}_{un}) ,
\label{eq:physical-cs}
\end{eqnarray}
where $d\sigma^{Ex}_{un}/dtd\phi_\pi$ refers to the experimentally observed cross section, $d\sigma^{ph,1\gamma}_{un}/dtd\phi_\pi$ refers to the physical cross section via OPE, and  $\delta^{ph,2\gamma}_{un}$ refers to the physical TPE correction to the cross section. Since actually we do not known all the dynamics of QCD, the physical $d\sigma^{ph,1\gamma}_{un}/dtd\phi_\pi$ and $\delta^{ph,2\gamma}_{un}$ are difficult to calculate  precisely.  It is a good approximation to assume $\delta^{ph,2\gamma}_{un} \approx \delta^{2\gamma}_{un}$ since the most important contributions in the OPE and TPE levels are considered in our calculation, respectively. We can expect that the model dependence of their ratio is much weaker than the absolute magnitude, like the $ep\rightarrow ep$ case where the relative TPE corrections are not sensitive to the input form factors. By this approximation, we have
\begin{eqnarray}
\frac{d\bar{\sigma}^{Ex}_{un}}{dtd\phi_\pi} \equiv  \frac{d\sigma^{ph,1\gamma}_{un}}{dtd\phi_\pi} \approx \frac{d\sigma^{Ex}_{un}}{dtd\phi_\pi}(1- \delta^{2\gamma}_{un}).
\label{eq:Subtracted-cs}
\end{eqnarray}

The current experimental analysis is based the experimental cross section $d\sigma^{Ex}_{un}/dtd\phi_\pi$ and Eq. (\ref{eq:OPE-cs-separared-form}). After considering the TPE contributions, in principle the analysis should be based on the corrected experimental cross section $d\bar{\sigma}^{Ex}_{un}/dtd\phi_\pi$ and Eq. (\ref{eq:OPE-cs-separared-form}). The comparison between the results from these two analyses can tell us how large are the TPE corrections to the separated cross sections $\sigma_{\textrm{L}}$, $\sigma_{\textrm{T}}$, $\sigma_{\textrm{LT}} $, and $ \sigma_{\textrm{TT}} $.

\begin{table}[ht]
\centering
\begin{tabular}{|p{2cm}<{\centering}|p{2cm}<{\centering}|p{2cm}<{\centering}|p{2cm}<{\centering}|p{2cm}<{\centering}|p{2cm}<{\centering}|
p{2cm}<{\centering}|}
\hline $Q^2\text{(GeV}^2)$ & $W\text{(GeV)}$ &   $t\text{(GeV}^2)$ & $ \sigma^{\textrm{ExA}}_{\textrm{L}} $& $ \sigma^{\textrm{ExA}}_{\textrm{T}} $  &$ \sigma^{\textrm{ExA}}_{\textrm{LT}} $ & $ \sigma^{\textrm{ExA}}_{\textrm{TT}} $\\
\hline  0.945 & 1.970  & -0.080  & 11.840 & 6.526    & 1.339   & -1.584\\
                            1.010 & 1.943  & -0.100  &  9.732 & 5.656    & 0.719   & -0.582\\
                            1.050 & 1.926  & -0.120  &  7.116 & 5.926    & 0.331   & -1.277\\
                            1.067 & 1.921  & -0.140  &  4.207 & 5.802    & 0.087   & -0.458 \\
\hline  1.532 & 1.975  & -0.165  &  4.378 & 3.507    & 0.356   & -0.268\\
                            1.610 & 1.944  & -0.195  &  3.191 & 3.528    & 0.143   & -0.126\\
                            1.664 & 1.924  & -0.225  &  2.357 & 2.354    &-0.028   & -0.241\\
                            1.702 & 1.911  & -0.255  &  2.563 & 2.542    &-0.100   & -0.083\\
\hline
\end{tabular}
\caption{Numerical results for the separated cross sections $ \sigma_{\textrm{L}}^{\textrm{ExA}}, \sigma_{\textrm{T}}^{\textrm{ExA}},  \sigma_{\textrm{LT}}^{\textrm{ExA}}, \sigma_{\textrm{TT}}^{\textrm{ExA}}$ directly taken from JLab $F_\pi$ \cite{electro-pion-production-JLab-2}.}
\label{table:ExA}
\end{table}

\begin{table}[ht]
\centering
\begin{tabular}{|p{2cm}<{\centering}|p{2cm}<{\centering}|p{2cm}<{\centering}|p{2cm}<{\centering}|p{2cm}<{\centering}|
p{2cm}<{\centering}|p{2cm}<{\centering}|}
\hline $Q^2\text{(GeV}^2)$ & $W\text{(GeV)}$ &   $t\text{(GeV}^2)$ & $\sigma^{\text{\textrm{ExB}}}_{\textrm{L}}$& $\sigma^{\text{\textrm{ExB}}}_{\textrm{T}}$  &$\sigma^{\text{\textrm{ExB}}}_{\textrm{LT}}$ & $\sigma^{\text{\textrm{ExB}}}_{\textrm{TT}}$\\
\hline  0.945 & 1.970  & -0.080  & 11.8344 & 6.8054    & 1.0266   & -0.8270 \\
                            1.010 & 1.943  & -0.100  &  8.4637 & 5.9616    & 0.5703   & -0.8978 \\
                            1.050 & 1.926  & -0.120  &  6.0577 & 5.3624    & 0.0985   & -1.0413 \\
                            1.067 & 1.921  & -0.140  &  4.2969 & 4.9353    &-0.0427   & -1.2690 \\
\hline  1.532 & 1.975  & -0.165  &  4.6398 & 3.7839    & 0.3938   & -0.1479 \\
                            1.610 & 1.944  & -0.195  &  3.3657 & 3.3617    & 0.1901   & -0.1395 \\
                            1.664 & 1.924  & -0.225  &  2.4370 & 3.0447    & 0.0299   & -0.1460 \\
                            1.702 & 1.911  & -0.255  &  1.7574 & 2.7978    &-0.0997   & -0.1569 \\
\hline
\end{tabular}
\caption{Numerical results for the separated cross sections $ \sigma_{\textrm{L}}^{\textrm{ExB}}, \sigma_{\textrm{T}}^{\textrm{ExB}},  \sigma_{\textrm{LT}}^{\textrm{ExB}}, \sigma_{\textrm{TT}}^{\textrm{ExB}}$ produced by the fitted formulas given in Ref \cite{electro-pion-production-JLab-2}.}
\label{table:ExB}
\end{table}

In the practical analysis, we take two data sets  named ExA and ExB as inputs to do the analysis.
In the data sets ExA, we take the experimental extracted $\sigma^{\textrm{ExA}}_{\textrm{L}}$, $\sigma^{\textrm{ExA}}_{\textrm{T}}$, $ \sigma^{\textrm{ExA}}_{\textrm{LT}}$, and $\sigma^{\textrm{ExA}}_{\textrm{TT}}$ by JLab $F_\pi$ \cite{electro-pion-production-JLab-2} as inputs to get $d\sigma^{\textrm{\textrm{ExA}}}_{un}/dtd\phi_\pi$ at specific $\epsilon$ and $\phi_\pi$ via Eq. (\ref{eq:OPE-cs-separared-form}). The corresponding values are listed in Table. \ref{table:ExA}.  We take $\epsilon$ as $0.33, 0.65$  at low $Q^2$, take $\epsilon$ as $ 0.27, 0.63$ at high $Q^2$, and take $\phi_{\pi}$ from $5^\circ$ to $355^\circ$ with $\Delta \phi_{\pi} = 25^\circ$.  In the data sets ExB, we use the experimentally fitted formula \cite{electro-pion-production-JLab-2} to produce  $d\sigma^{\textrm{\textrm{ExB}}}_{un}/dtd\phi_\pi$. For comparison, the corresponding $ \sigma^{\textrm{ExB}}_{\textrm{L}}$, $\sigma^{\textrm{ExB}}_{\textrm{T}}$, $ \sigma^{\textrm{ExB}}_{\textrm{LT}}$, and $ \sigma^{\textrm{ExB}}_{\textrm{TT}}$ are listed in Table. \ref{table:ExB}. In this data sets, we take $\epsilon$ from 0.33 to 0.65 with $\Delta \epsilon=0.03$ at low $Q^2$, take $\epsilon$ from 0.27 to 0.63 with $\Delta \epsilon = 0.035$ at high $Q^2$ to produce more data points, and take  $\phi_{\pi}$ from $5^\circ$ to $355^\circ$ with $\Delta \phi_{\pi} = 25^\circ$.

After getting the data sets $d\sigma^{\textrm{\textrm{ExA,ExB}}}_{un}/dtd\phi_\pi$, we use the estimated TPE corrections in the corresponding kinematics region to get $d\bar{\sigma}^{\textrm{\textrm{ExA,ExB}}}_{un}/dtd\phi_\pi$. Then we use Eq.  (\ref{eq:OPE-cs-separared-form}) to fit the corrected data sets  to get the corrected separated cross sections $ \bar{\sigma}^{\textrm{ExA,ExB}}_{\textrm{L}}$, $\bar{\sigma}^{\textrm{ExA,ExB}}_{\textrm{T}}$, $ \bar{\sigma}^{\textrm{ExA,ExB}}_{\textrm{LT}}$, and $\bar{\sigma}^{\textrm{ExA,ExB}}_{\textrm{TT}}$.

\begin{table}[ht]
\centering
\begin{tabular}{|p{2cm}<{\centering}|p{2cm}<{\centering}|p{2cm}<{\centering}|p{2cm}<{\centering}|p{2cm}<{\centering}|p{2cm}<{\centering}|
p{2cm}<{\centering}|}
\hline $Q^2\text{(GeV}^2)$ & $W\text{(GeV)}$ &   $t\text{(GeV}^2)$ & $\sigma^{\textrm{ExA}}_{\textrm{L}}$& ${\bar{\sigma}}^{\textrm{ExA}}_{\textrm{L}}/\sigma^{\textrm{ExA}}_{\textrm{L}}$  &$\sigma^{\textrm{ExB}}_{\textrm{L}}$ & ${\bar{\sigma}}^{\textrm{ExB}}_{\textrm{L}}/\sigma^{\textrm{ExB}}_{\textrm{L}}$\\
\hline                      0.945 & 1.970  & -0.080   & 11.840 & 0.9209    & 11.8344   & 0.9191\\
                            1.010 & 1.943  & -0.100   & 9.732   & 0.9137    & 8.4637   & 0.8967\\
                            1.050 & 1.926  & -0.120   & 7.116   & 0.8726    & 6.0577   & 0.8664\\
                            1.067 & 1.921  & -0.140   & 4.207   & 0.7820    & 4.2969   & 0.8243\\
\hline                      1.532 & 1.975  & -0.165   & 4.378  & 0.8518    &4.6398     & 0.8510\\
                            1.610 & 1.944  & -0.195   & 3.191   & 0.7839    &3.3657    & 0.8083\\
                            1.664 & 1.924  & -0.225   & 2.357   & 0.7095    &2.4370    & 0.7490\\
                            1.702 & 1.911  & -0.255   & 2.563   & 0.7946    &1.7574    & 0.6669\\
\hline
\end{tabular}
\caption{Numerical results for the ratios ${\bar{\sigma}}^{\textrm{ExA}}_{\textrm{L}}/\sigma^{\textrm{ExA}}_{\textrm{L}}$ and ${\bar{\sigma}}^{\textrm{ExB}}_{\textrm{L}}/\sigma^{\textrm{ExB}}_{\textrm{L}}$, the experimental data sets for $\sigma^{\textrm{ExA}}_{\textrm{L}}$ and $\sigma^{\textrm{ExB}}_{\textrm{L}}$ are also listed.}
\label{table:TPE-sigma-L}
\end{table}

\begin{table}[ht]
\centering
\begin{tabular}{|p{2cm}<{\centering}|p{2cm}<{\centering}|p{2cm}<{\centering}|p{2cm}<{\centering}|p{2cm}<{\centering}|p{2cm}<{\centering}|
p{2cm}<{\centering}|}
\hline $Q^2\text{(GeV}^2)$ & $W\text{(GeV)}$ &   $t\text{(GeV}^2)$ & $\sigma^{\textrm{ExA}}_{\textrm{T}}$& ${\bar{\sigma}}^{\textrm{ExA}}_{\textrm{T}}/\sigma^{\textrm{ExA}}_{\textrm{T}}$  &$\sigma^{\textrm{ExB}}_{\textrm{T}}$ & ${\bar{\sigma}}^{\textrm{ExB}}_{\textrm{T}}/\sigma^{\textrm{ExB}}_{\textrm{T}}$\\
\hline                      0.945 & 1.970  & -0.080  & 6.526 & 1.2064    & 6.8504   & 1.2027\\
                            1.010 & 1.943  & -0.100  & 5.656 & 1.2104    & 5.9616   & 1.1990\\
                            1.050 & 1.926  & -0.120  & 5.926 & 1.1977    & 5.3624   & 1.1932\\
                            1.067 & 1.921  & -0.140  & 5.802 & 1.1858    & 4.9353   & 1.1853\\
\hline                      1.532 & 1.975  & -0.165  & 3.507 & 1.2312    & 3.7839   & 1.2278\\
                            1.610 & 1.944  & -0.195  & 3.528 & 1.2273    & 3.3617   & 1.2275\\
                            1.664 & 1.924  & -0.225  & 2.354 & 1.2269    & 3.0447   & 1.2271\\
                            1.702 & 1.911  & -0.255  & 2.542 & 1.2453    & 2.7978   & 1.2267\\
\hline
\end{tabular}
\caption{Numerical results for the ratios ${\bar{\sigma}}^{\textrm{ExA}}_{\textrm{T}}/\sigma^{\textrm{ExA}}_{\textrm{T}}$ and ${\bar{\sigma}}^{\textrm{ExB}}_{\textrm{T}}/\sigma^{\textrm{ExB}}_{\textrm{T}}$, the experimental data sets for $\sigma^{\textrm{ExA}}_{\textrm{T}}$ and $\sigma^{\textrm{ExB}}_{\textrm{T}}$ are also listed.}
\label{table:TPE-TPE-sigma-T}
\end{table}

\begin{table}[ht]
\centering
\begin{tabular}{|p{2cm}<{\centering}|p{2cm}<{\centering}|p{2cm}<{\centering}|p{2cm}<{\centering}|p{2cm}<{\centering}|p{2cm}<{\centering}|
p{2cm}<{\centering}|}
\hline $Q^2\text{(GeV}^2)$ & $W\text{(GeV)}$ &   $t\text{(GeV}^2)$ & $\sigma^{\textrm{ExA}}_{\textrm{LT}}$& ${\bar{\sigma}}^{\textrm{ExA}}_{\textrm{LT}}/\sigma^{\textrm{ExA}}_{\textrm{LT}}$  &$\sigma^{\textrm{ExB}}_{\textrm{LT}}$ & ${\bar{\sigma}}^{\textrm{ExB}}_{\textrm{LT}}/\sigma^{\textrm{ExB}}_{\textrm{LT}}$\\
\hline                      0.945 & 1.970  & -0.080  & 1.339 & 1.2046    & 1.0266   & 1.2730\\
                            1.010 & 1.943  & -0.100  & 0.719 & 1.3394    & 0.5703   & 1.4263\\
                            1.050 & 1.926  & -0.120  & 0.331 & 1.7029    & 0.0985   & 3.0947\\
                            1.067 & 1.921  & -0.140  & 0.087 & 3.5629    &-0.0427   & 0.5850\\
\hline                      1.532 & 1.975  & -0.165  & 0.356 & 1.4101    & 0.3938   & 1.4284\\
                            1.610 & 1.944  & -0.195  & 0.143 & 1.9492    & 0.1901   & 1.7674\\
                            1.664 & 1.924  & -0.225  &-0.028 &-3.5338    & 0.0299   & 5.4036\\
                            1.702 & 1.911  & -0.255  &-0.100 &-0.1846    &-0.0997   &-0.2142\\
\hline
\end{tabular}
\caption{Numerical results for the ratios ${\bar{\sigma}}^{\textrm{ExA}}_{\textrm{LT}}/\sigma^{\textrm{ExA}}_{\textrm{LT}}$ and ${\bar{\sigma}}^{\textrm{ExB}}_{\textrm{LT}}/\sigma^{\textrm{ExB}}_{\textrm{LT}}$,  the experimental data sets for $\sigma^{\textrm{ExA}}_{\textrm{LT}}$ and $\sigma^{\textrm{ExB}}_{\textrm{LT}}$ are also listed.}
\label{table:TPE-sigma-LT}
\end{table}

\begin{table}[ht]
\centering
\begin{tabular}{|p{2cm}<{\centering}|p{2cm}<{\centering}|p{2cm}<{\centering}|p{2cm}<{\centering}|p{2cm}<{\centering}|p{2cm}<{\centering}|
p{2cm}<{\centering}|}
\hline $Q^2\text{(GeV}^2)$ & $W\text{(GeV)}$ &   $t\text{(GeV}^2)$ & $\sigma^{\textrm{ExA}}_{\textrm{TT}}$& $\bar{\sigma}^{\textrm{ExA}}_{\textrm{TT}}/\sigma^{\textrm{ExA}}_{\textrm{TT}}$  &$\sigma^{\textrm{ExB}}_{\textrm{TT}}$ & $\bar{\sigma}^{\textrm{ExB}}_{\textrm{TT}}/\sigma^{\textrm{ExB}}_{\textrm{TT}}$\\
\hline                      0.945 & 1.970  & -0.080  &-1.584 & 1.0024    &-0.8270   & 0.9470\\
                            1.010 & 1.943  & -0.100  &-0.582 & 0.8963    &-0.8978   & 0.9554\\
                            1.050 & 1.926  & -0.120  &-1.277 & 0.9777    &-1.0413   & 0.9711\\
                            1.067 & 1.921  & -0.140  &-0.458 & 0.7903    &-1.2690   & 0.9917\\
\hline                      1.532 & 1.975  & -0.165  &-0.268 & 0.8833    &-0.1479   & 0.6491\\
                            1.610 & 1.944  & -0.195  &-0.126 & 0.6490    &-0.1395   & 0.6349\\
                            1.664 & 1.924  & -0.225  &-0.241 & 0.8344    &-0.1460   & 0.6333\\
                            1.702 & 1.911  & -0.255  &-0.083 & 0.3141    &-0.1569   & 0.6378\\
\hline
\end{tabular}
\caption{Numerical results for the ratios $\bar{\sigma}^{\textrm{ExA}}_{\textrm{TT}}/\sigma^{\textrm{ExA}}_{\textrm{TT}}$ and $\bar{\sigma}^{\textrm{ExB}}_{\textrm{TT}}/\sigma^{\textrm{ExB}}_{\textrm{TT}}$, the experimental data sets for $\sigma^{\textrm{ExA}}_{\textrm{TT}}$ and $\sigma^{\textrm{ExB}}_{\textrm{TT}}$ are also listed.}
\label{table:TPE-sigma-TT}
\end{table}

In Tables \ref{table:TPE-sigma-L}-\ref{table:TPE-sigma-TT}, we present the relative TPE corrections $\bar{\sigma}^{\textrm{ExA}}_{\textrm{X}}/\sigma^{\textrm{ExA}}_{\textrm{X}}$ and $\bar{\sigma}^{\textrm{ExB}}_{\textrm{X}}/\sigma^{\textrm{ExB}}_{\textrm{X}}$ where $\textrm{X}$ refers to $\textrm{L}, \textrm{T}, \textrm{LT}$ and $\textrm{TT}$, respectively. The numerical results show a general property that both data sets give similar relative TPE corrections to $ \sigma_{\textrm{L,T,TT}}^{\textrm{ExA,ExB}}$ and give very different relative TPE corrections to $ \sigma_{\textrm{LT}}^{\textrm{ExA,ExB}}$ at some special points. The latter can be understood in a simple way since, in these points, the input data sets $\sigma_{\textrm{LT}}^{\textrm{ExA,ExB}}$ are much smaller than the others. This means that the relative uncertainty to the extracted $\sigma_{\textrm{LT}}^{\textrm{ExA,ExB}}$ actually is much larger than others.

At $Q^2\approx 1$ GeV $^2$ and $-t\approx 0.1$ GeV$^2$, the relative TPE corrections to $\sigma_{\textrm{L}}^{\textrm{ExA,ExB}}$ are about $-10\%$ and the corrections to $\sigma_{\textrm{T}}^{\textrm{ExA,ExB}}$ are about $20\%$. When $Q^2$ and $-t$ increase, the relative TPE corrections to $\sigma_{\textrm{L}}^{\textrm{ExA,ExB}}$ reach about from $-20\%$ to $-30\%$, while are still about $20\%$ to $\sigma_{\textrm{T}}^{\textrm{ExA,ExB}}$.  The relative TPE corrections to $\sigma_{\textrm{TT}}^{\textrm{ExA,ExB}}$ are small at small $-t$ and are large and sensitive to the input data sets at large $-t$. The TPE corrections to $\sigma_{\textrm{LT}}^{\textrm{ExA,ExB}}$ are always large and even become un-reliable and very sensitive to the input data sets at large $-t$. The experimentally extracted $\sigma_{\textrm{L}}$  is usually used to determine the pion form factor $F_{\pi}$ through the Chew-Low method (based on the Born term model \cite{Born-term-model}) or  the Regge model \cite{VGL-model}. Our results show that the relative TPE corrections to $\sigma_{\textrm{L}}^{\textrm{ExA,ExB}}$ reach about from $-10\%$ to $-30\%$ at $Q^2=1$--$1.6$ GeV$^2$. This means that the relative TPE corrections to the EM form factor of pions are about on the order of $-5\%$ to $-15\%$ and should be considered carefully. At high $Q^2$, one can expect that the TPE corrections should be much more important and should be considered seriously to extract the EM form factor of pions reliablely.

In summary, in this work the TPE corrections to the amplitude and the unpolarized differential cross section of $ep\rightarrow en\pi^+$ are estimated in a hadronic model. The TPE corrections to the extracted  four separated cross sections are also analyzed based on the experimental data sets. Our results show that at $Q^2=1$--$1.6$ GeV$^2$, the TPE correction to $\sigma_{\textrm{L}}$ is about from $-10\%$ to $-30\%$ and about $20\%$ to $\sigma_{\textrm{T}}$.

\section{Acknowledgments}

The author Hai-Qing Zhou would like to thank Zhi-Yong Zhou and Dian-Yong Chen for their kind and helpful discussions.  This work was supported by the  National Natural Science Foundations of China under Grants No. 11375044 and No. 11975075.

\section{Appendix: The momenta in the Lab frame and center frame of $n\pi^+$}

In this appendix, we list the manifest expressions of the momenta used in our calculation in the laboratory frame and the center frame of $n\pi^+$. In the center frame of pions ($p_4$) and neutrons ($p_5$), the momenta labeled as $p_{iC}$ are taken as
\begin{eqnarray}
p_{1C} &=& ( E_{1C}, E_{1C} sin\theta_1,  0 , E_{1C} cos\theta_1 ), \nonumber\\
p_{2C} &=& ( E_{2C}, 0, 0 ,  -\sqrt{E^2_{2C}-M_n^2}  ), \nonumber\\
p_{45C} &\equiv& p_{4C}+p_{5C}=( W, 0, 0,0 ), \nonumber\\
q_{C}  &\equiv& p_{45C}-p_{2C}= (W-E_{2C}, 0, 0, \sqrt{E_{2C}^{2}-M_n^2}), \nonumber\\
p_{3C} &=& p_{1C}-q_{C}, \nonumber\\
p_{4C} &=& ( E_{\pi C},\ p_{\pi C} sin\theta_{\pi} cos\phi_{\pi}, p_{\pi C}\sin\theta_{\pi} \sin\phi_{\pi}, p_{\pi C}\cos\theta_{\pi} ), \nonumber\\
p_{5C} &=& p_{45C}-p_{4C}.
\end{eqnarray}

In the laboratory frame the momenta labeled as $p_{iL}$ are taken as
\begin{eqnarray}
p_{1L} &=& (E_{e},0, 0 ,E_{e}), \nonumber\\
p_{2L} &=& (m_p, 0, 0 , 0 ), \nonumber\\
p_{3L} &=& (E_{e'},E_{e'} sin\theta_{e'}, 0, E_{e'} cos\theta_{e'}  ).
\end{eqnarray}
From these expressions, we have the following relations.
\begin{eqnarray}\label{equ:t-and-p14}
s &=& \frac{1}{2}\Big[\ {m_p}^2+Q^2+W^2+\frac{\sqrt{(({m_p}^2+Q^2)^2+2(-m_p^2+Q^2)W^2+W^4) (1-\epsilon^2)}}{1-\epsilon}\ \Big],\nonumber\\
t &=& \frac{1}{2W^2} \Big[ -m_p^4+(m_{\pi}^2-W^2)(Q^2+W^2)+m_p^2(m_{\pi}^2-Q^2+2W^2)\nonumber\\
&& + \ t_0 \sqrt{m_p^4+2 m_p^2(Q^2-W^2)+(Q^2+W^2)^2} \cos\theta_{\pi} \Big], \nonumber\\
p_{14} &=& \frac{1}{4W^2}\Big\{\  \frac{ t_0 [-(m_p^2+Q^2)t_1+W^2 t_2]\cos\theta_{\pi} }{\sqrt{m_p^4+2m_p^2(Q^2-W^2)+(Q^2+W^2)^2}} + t_1(m_p^2-m_{\pi}^2-W^2) \nonumber\\
&& -   2W t_0 \  \sqrt{\frac{-Q^2 t_1 s+Q^2(m_p^2-s)W^2}{[(m_p^2+Q^2)^2+2(-m_p^2+Q^2)W^2+W^4]}}\sin{\theta_{\pi}}\cos{\phi_{\pi}} \Big\}, \end{eqnarray}
with
\begin{eqnarray}
t_0 &=& \sqrt{(m_p-m_{\pi}-W)(m_p+m_{\pi}-W)(m_p-m_{\pi}+W)(m_p+m_{\pi}+W)}, \nonumber\\
t_1 &=& m_p^2+Q^2-s, \nonumber\\
t_2 &=& m_p^2-Q^2-s.
\end{eqnarray}
and
\begin{eqnarray}\label{equ:epsilon}
\epsilon &\equiv& \Big[\ 1+ \frac{m_p^4+2m_p^2(Q^2-W^2)+(Q^2+W^2)^2}{2m_p^2 Q^2 } \tan^2{\frac{\theta_e^{\prime}}{2}} \Big]^{-1}.
\end{eqnarray}

The expressions of the kinematics are consistent with those used in the JLab $F_\pi$ experiment \cite{electro-pion-production-JLab-2}.


\begin{thebibliography}{99}

\bibitem{hadronic model}
P. G. Blunden, W. Melnitchuk, and J. A. Tjon, Phys. Rev.Lett. {\bf 91}, 142304 (2003);
S. Kondratyuk, P. G. Blunden, W. Melnitchuk, and J. A.Tjon, ibid {\bf 95}, 172503 (2005);
P. G. Blunden, W. Melnitchuk, and J. A. Tjon, Phys. Rev. C {\bf 72}, 034612 (2005).

\bibitem{GPD method}
Y. C. Chen, A. Afanasev, S. J. Brodsky, C. E. Carlson, and M. Vanderhaeghen, Phys. Rev. Lett. {\bf 93}, 122301 (2004);
A. Afanasev, S. J. Brodsky, C. E. Carlson, Y. C. Chen, and M. Vanderhaeghen, Phys. Rev. D {\bf 72}, 013008 (2005).

\bibitem{pQCD method}
N. Kivel and M. Vanderhaeghen, Phys. Rev. Lett. {\bf 103}, 092004 (2009);
D. Borisyuk and A. Kobushkin, Phys. Rev. C {\bf 79}, 034001 (2009).



\bibitem{dispersion relation-1}
D. Borisyuk and A. Kobushkin, Phys. Rev. C {\bf 74}, 065203 (2006);
D. Borisyuk and A. Kobushkin, Phys. Rev. C {\bf 83}, 025203 (2011);
D. Borisyuk and A. Kobushkin, Phys. Rev. C {\bf 86}, 055204 (2012);
D. Borisyuk and A. Kobushkin, Phys. Rev. C {\bf 89}, 025204 (2014);

\bibitem{dispersion relation-2}
O. Tomalak and M. Vanderhaeghen, Eur. Phys. J. A {\bf 51}, 24 (2015);
P. G. Blunden and W. Melnitchouk, Phys. Rev. C {\bf 95}, 065209 (2017).
Oleksandr Tomalak, Barbara Pasquini, and Marc Vanderhaeghen, Phys. Rev. D {\bf96},096001 (2017)


\bibitem{SCEF}
N. Kivel and M. Vanderhaeghen, J. High Energy Phys. {\bf 04}, 029 (2013).

\bibitem{phenomenological parametrizations}
Y. C. Chen, C. W. Kao, and S. N. Yang, Phys. Lett. B {\bf 652}, 269 (2007);
D. Borisyuk and A. Kobushkin, Phys. Rev. C {\bf 76}, 022201 (2007).

\bibitem{TPE-ee-ppbar}
D. Y. Chen, H. Q. Zhou, Y.B. Dong, Phys. Rev. C {\bf 78}, 045208 (2008).

\bibitem{TPE-epi-epi}
D. Borisyuk and A. Kobushkin, Phys. Rev. C {\bf 78}, 025208 (2008);
P.G. Blunden, W. Melnitchouk, and J. A. Tjon, Phys. Rev. C {\bf 81}, 018202 (2010);
Yu Bing Dong and S. D. Wanga, Phys. Lett. B {\bf 684}, 123 (2010).

\bibitem{Afanasev2013}
Andrei Afanasev, Aleksandrs Aleksejevs and Svetlana Barkanova, Phys. Rev. C {\bf 88},053008 (2013).

\bibitem{TPE-mup-mup}
Dian-Yong Chen and Yu-Bing Dong, Phys. Rev. C {\bf 87}, 045209 (2013);
O. Tomalak and M. Vanderhaeghen, Phys. Rev. D {\bf 90}, 013006 (2014);
O. Koshchii and A. Afanasev, Phys. Rev. D {\bf 94}, 116007 (2016);
Hai-Qing Zhou, Phys. Rev. C {\bf 95}, 025203,(2017);
Oleksandr Koshchii and Andrei Afanasev, Phys.Rev. D {\bf100},096020 (2019).

\bibitem{TPE-ep-eppi0}
Hai-Qing Zhou and Shin Nan Yang, Phys. Rev. C {\bf 96}, 055210 (2017).


\bibitem{electro-pion-production-Cornell}
C. J. Bebek \textit{et al}., Phys. Rev. D {\bf 13}, 25 (1976);
C. J. Bebek \textit{et al}., Phys. Rev. D {\bf 17}, 1693 (1978).

\bibitem{electro-pion-production-DESY}
P. Brauel \textit{et al}. (DESY), Phys. Lett. B {\bf 65}, 184 (1976);
P. Brauel \textit{et al}., Phys. Lett. B {\bf 69}, 253 (1977);
H. Ackermann \textit{et a}l., Nucl. Phys. B {\bf 137}, 294 (1978);
P. Brauel \textit{et al}., Z. Phys. C {\bf 3}, 101 (1979).


\bibitem{electro-pion-production-JLab-1}
J. Volmer \textit{et al}. (Jefferson Lab F$\pi$ Collaboration), Phys. Rev. Lett. {\bf 86}, 1713 (2001);
T. Horn \textit{et a}l. (Jefferson Lab F$\pi$ Collaboration), Phys. Rev. lett. {\bf 97}, 192001 (2006);
V. Tadevosyan \textit{et al}. (Jefferson Lab F$\pi$ Collaboration), Phys. Rev. C {\bf 75}, 055205 (2007).

\bibitem{electro-pion-production-JLab-2}
H. P. Blok, T. Horn \textit{et al}. (Jefferson Lab F$\pi$ Collaboration), Phys. Rev. C {\bf 78}, 045202 (2008).


\bibitem{electro-pion-production-JLab-3}
G.M. Huber,\textit{ et al}. (Jefferson Lab F$\pi$ Collaboration), Phys. Rev. C {\bf 78}, 045203 (2008).



\bibitem{zhouhq2011-pion-photon-interaction}
Hai Qing Zhou, Phys. Lett. B {\bf 706}, 82-85, (2011).

\bibitem{Pacakge X}
H. H. Patel,  Comput. Phys. Commun. {\bf 197}, 276-290, (2015).

\bibitem{IR-Mo-and-Tsai}
L. W. Mo and Y. S. Tsai, Rev. Mod. Phys. {\bf 41}, 205 (1969);
Y. S. Tsai, Phys. Rev. {\bf 122}, 1898 (1961).

\bibitem{IR-Maximon-and-Tjon}
L. C. Maximon and J. A. Tjon, Phys. Rev. C {\bf 62}, 054320 (2000).



\bibitem{Blunden2010-pion-form-factor}
P. G. Blunden, W. Melnitchouk, J. A. Tjon, Phys. Rev. C {\bf 81}, 018202 (2010).

\bibitem{FenyCalc}
Vladyslav Shtabovenko, Rolf Mertig and Frederik Orellana, Comput. Phys. Commun. {\bf 207}, 432 (2016);
R. Mertig, M. Bohm and Ansgar Denner, Comput. Phys. Commun. {\bf 64}, 345 (1991).

\bibitem{LoopTools}
T. Hahn, M. Perez-Victoria, Comput. Phys. Commun. {\bf 118}, 153 (1999).


\bibitem{Born-term-model}
A. Actor, J. G. Korner, and I. Bender, Nuovo Cim.  {\bf A} 24, 369 (1974).

\bibitem{VGL-model}
M. Vanderhaeghen, M. Guidal and J.-M. Laget, Nucl. Phys. A {\bf 627}, 645 (1997); Phys. Rev. C {\bf 57}, 1454 (1998).



\end{thebibliography}
\end{document}